\documentclass[a4paper,11pt]{article}
\pdfoutput=1 

\usepackage{jcappub} 

\usepackage{color}

\def\nat{Nature}

\def\mnras{Monthly Notices of the Royal Astronomical Society}

\def\apj{Astroph. J.}
\def\apjl{Astroph. J.}
\def\jcap{JCAP}
\def\aap{A\&A}

\def\pasj{PASJ}

\def\mnras{MNRAS}

\title{Electromagnetic emission of white dwarf binary mergers}

\author[a,b,c]{J.~A.~Rueda,}
\author[a,b,c,d]{R.~Ruffini,}
\author[a,b]{Y.~Wang,}
\author[a,b]{C.~L.~Bianco,}
\author[e,f]{J.~M.~Blanco-Iglesias,}
\author[a,b,d]{M.~Karlica,}
\author[e,f]{P.~Lor\'en-Aguilar,}
\author[a,b]{R.~Moradi}
\author[g]{and N.~Sahakyan}

\affiliation[a]{ICRA, Dipartimento di Fisica, 
                     Sapienza Universit\`a di Roma, 
                     P.le Aldo Moro 5, 
                     I--00185 Rome, 
                     Italy}
                     
\affiliation[b]{ICRANet, 
                     P.zza della Repubblica 10, 
                     I--65122 Pescara, 
                     Italy}

\affiliation[c]{ICRANet-Rio, 
                     Centro Brasileiro de Pesquisas F\'isicas, 
                     Rua Dr. Xavier Sigaud 150, 
                     22290--180 Rio de Janeiro, 
                     Brazil}
                     
\affiliation[d]{Universit\'e de Nice Sophia Antipolis, 
                     CEDEX 2, Grand Ch\^{a}teau Parc Valrose, 
                     Nice, 
                     France}

\affiliation[e]{Departament de F\'{i}sica, Universitat Polit\`{e}cnica de Catalunya, c/Esteve Terrades, 5, 08860 Castelldefels, Spain}

\affiliation[f]{School of Physics, University of Exeter, Stocker Road, Exeter EX4 4QL, UK}

\affiliation[g]{ICRANet-Armenia, Marshall Baghramian Avenue 24a, 0019 Yerevan, Armenia}

\emailAdd{jorge.rueda@icra.it}
\emailAdd{ruffini@icra.it}

\date{\today}

\abstract{
It has been recently proposed that the ejected matter from white dwarf (WD) binary mergers can produce transient, optical and infrared emission similar to the ``kilonovae'' of neutron star (NS) binary mergers. To confirm this we calculate the electromagnetic emission from WD-WD mergers and compare with kilonova observations. We simulate WD-WD mergers leading to a massive, fast rotating, highly magnetized WD with an adapted version of the smoothed-particle-hydrodynamics (SPH) code Phantom. We thus obtain initial conditions for the ejecta such as escape velocity, mass and initial position and distribution. The subsequent thermal and dynamical evolution of the ejecta is obtained by integrating the energy-conservation equation accounting for expansion cooling and a heating source given by the fallback accretion onto the newly-formed WD and its magneto-dipole radiation. 
We show that magnetospheric processes in the merger can lead to a prompt, short gamma-ray emission of up to $\approx 10^{46}$~erg in a timescale of $0.1$--$1$~s. The bulk of the ejecta initially expands non-relativistically with velocity $0.01~c$ and then it accelerates to $0.1~c$ due to the injection of fallback accretion energy. The ejecta become transparent at optical wavelengths around $\sim 7$~days post-merger with a luminosity $10^{41}$--$10^{42}$~erg~s$^{-1}$. The X-ray emission from the fallback accretion becomes visible around $\sim 150$--$200$~day post-merger with a luminosity of $10^{39}$~erg~s$^{-1}$. 
We also predict the post-merger time at which the central WD should appear as a pulsar depending on the value of the magnetic field and rotation period.
}

\begin{document}
\maketitle
\flushbottom

\section{Introduction}\label{sec:1}

It was recently shown in \citet{2018JCAP...10..006R} that WD-WD mergers can produce optical and infrared emission that resemble the one emitted from the ``kilonovae'' produced by NS-NS mergers. This novel, previously not addressed possibility of WD-WD mergers emission was there applied to the analysis of the optical and infrared observations of the ``kilonova'' AT 2017gfo \citep{2017Natur.551...67P,2017Natur.551...64A,2017ApJ...848L..17C,2017ApJ...848L..18N}, associated with GRB 170817A \citep{2017ApJ...848L..13A,2017ApJ...848L..14G}.

The emission in the optical and infrared wavelengths is of thermal character being due to the adiabatic cooling of WD-WD merger ejecta, which is also powered by the fallback accretion onto the newly-formed WD. The ejecta mass is about $10^{-3}~M_\odot$ \citep{2009A&A...500.1193L,2011ApJ...737...89D} and the fallback may inject $10^{47}$--$10^{49}$~erg~s$^{-1}$ at early times and then fall-off following a power-law behavior \citep{2009A&A...500.1193L}.

The thermal ejecta start to become transparent in the optical wavelengths at $t\sim 7$~days with a peak bolometric luminosity $L_{\rm bol}\sim 10^{42}$~erg~s$^{-1}$. These ejecta are therefore powered by a different mechanism with respect to the one in the kilonova from NS-NS which are powered by the radioactive decay of r-process heavy material synthesized in the merger.

Since the observational features of WD-WD mergers are an important topic by their own, the aim of this article is to give details on their expected electromagnetic emission, not only in the optical and infrared but also in the X- and gamma-rays. 

The article is organized as follows. In Sec.~\ref{sec:2} we recall the properties of the WD-WD mergers obtained from numerical simulations, Sec.~\ref{sec:3} is devoted to the analysis of the optical and infrared emission from the cooling of the merger ejecta. We show in Sec.~\ref{sec:5} the X-ray emission from fallback accretion and spindown of the newly-formed central WD, in Sec.~\ref{sec:6} we present a brief discussion on the possible prompt emission in gamma-rays, and in Sec.~\ref{sec:7} we present the summary and the conclusions of the article.

\section{WD-WD mergers}\label{sec:2}

\subsection{Post-merger configuration}

Numerical simulations of WD-WD mergers indicate that, when the merger does not lead to a prompt type Ia supernova (SN) explosion, the  merged configuration has in general three distinct regions \citep{1990ApJ...348..647B, 2004A&A...413..257G, 2009A&A...500.1193L,2012A&A...542A.117L,2012ApJ...746...62R,
2013ApJ...767..164Z, 2014MNRAS.438...14D}: a rigidly rotating, central WD, on top of which there is a hot, differentially-rotating, convective corona, surrounded by a rapidly rotating Keplerian disk. The corona is composed of  about half of the  mass of the secondary star which is totally disrupted and roughly the other half of the secondary mass is in the disk. Little mass ($\sim 10^{-3}\,M_\odot$) is ejected during the merger.

Depending on the merging component masses, the central remnant can be a massive ($1.0$--$1.5~M_\odot$), highly magnetized ($10^9$--$10^{10}$~G) and fast rotating ($P=1$--$10$~s) WD \cite{2013ApJ...772L..24R,2018ApJ...857..134B}.

\begin{figure*}
\centering
\includegraphics[width=0.32\hsize,clip]{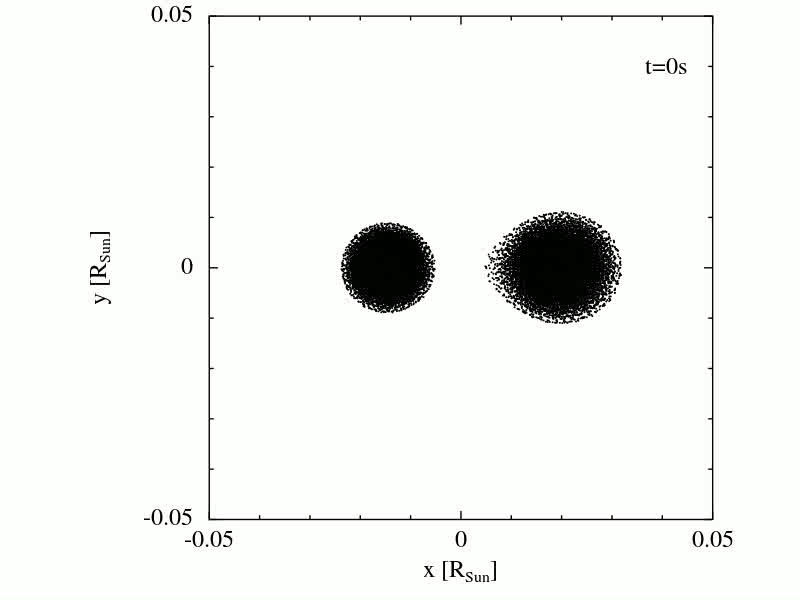}
\includegraphics[width=0.32\hsize,clip]{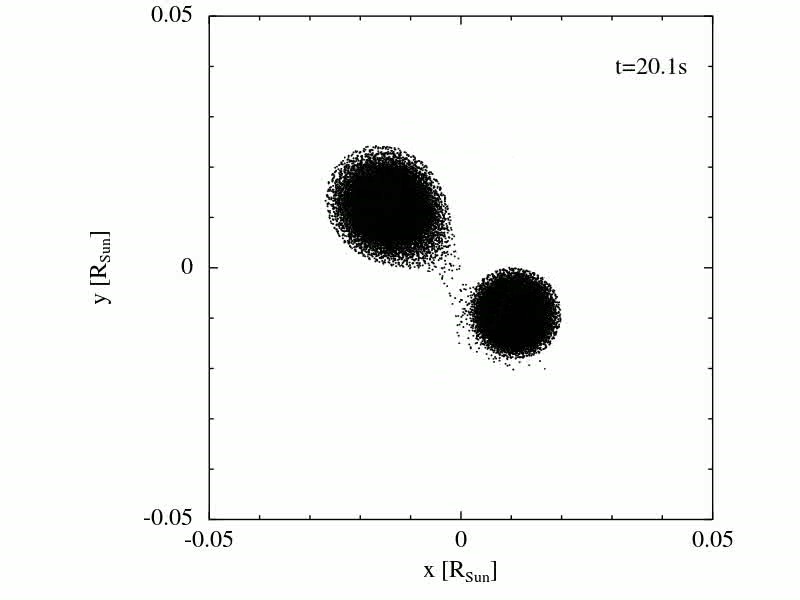}
\includegraphics[width=0.32\hsize,clip]{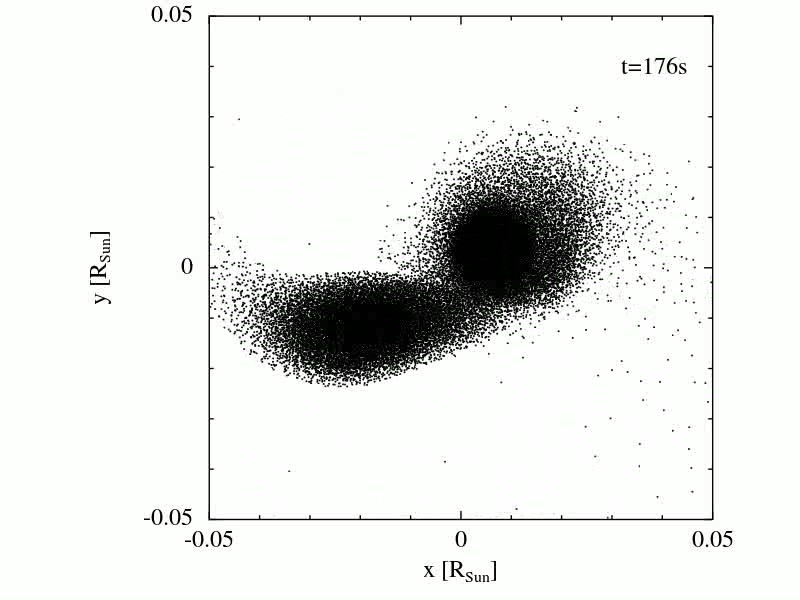}
\includegraphics[width=0.32\hsize,clip]{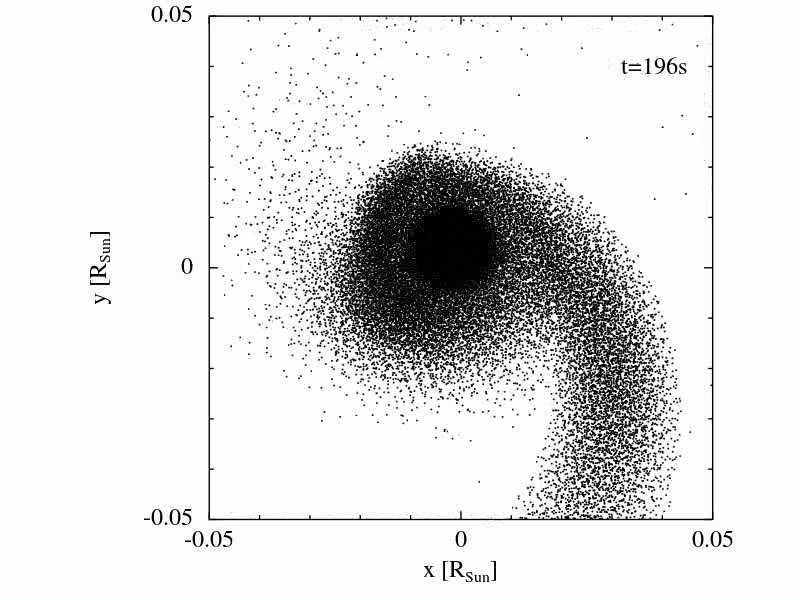}
\includegraphics[width=0.32\hsize,clip]{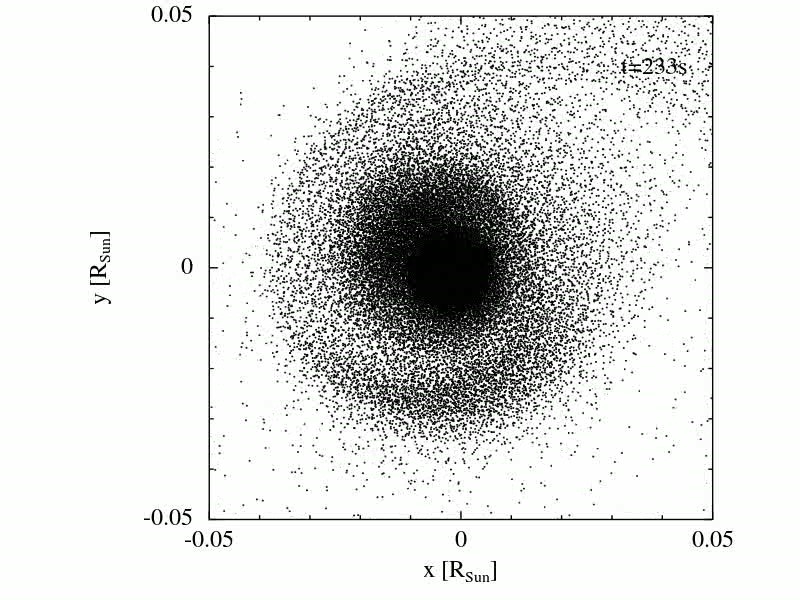}
\includegraphics[width=0.32\hsize,clip]{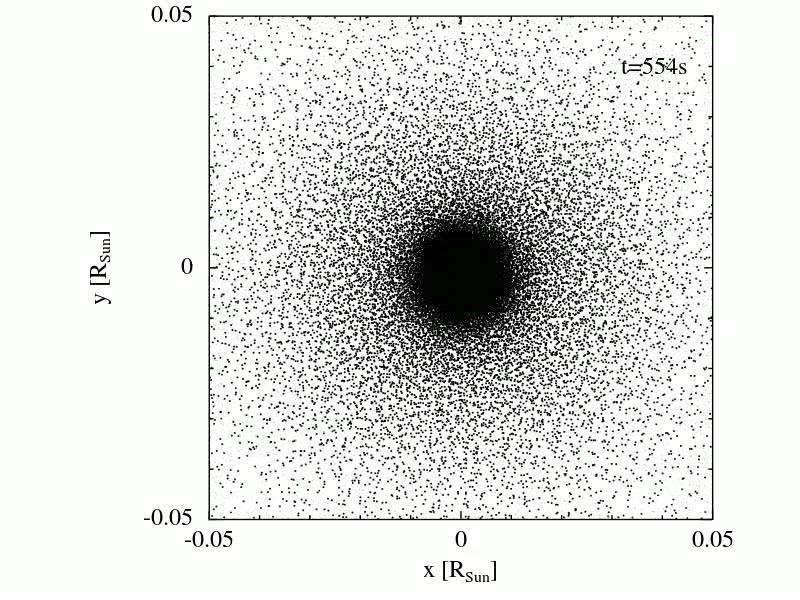}
\caption{Snapshots of the time evolution of $0.8+0.6~M_\odot$ WD-WD merger from the SPH simulation with $7\times 10^4$ particles. The newly-formed central WD has approximately $1.1~M_\odot$. In the sequence it can be seen how the secondary star is disrupted by Roche lobe overflow. Nearly half of the mass of the secondary star is transferred to the primary and the rest remains bound to the newly-formed central WD in form of a Keplerian disk. Little mass is ejected, in the present simulation nearly $1.2\times 10^{-3}~M_\odot.$}\label{fig:WDWDmerger}
\end{figure*}

Figure~\ref{fig:WDWDmerger} shows a series of snapshots of the time evolution of a $0.8+0.6~M_\odot$ WD-WD merger obtained by an adapted version of the smoothed-particle-hydrodynamics (SPH) code Phantom \citep{2017arXiv170203930P,2017ascl.soft09002P}. This simulation was run with $7\times 10^4$ SPH particles. The newly-formed central WD has approximately $1.1~M_\odot$. The ejected mass has been estimated to be $1.2\times 10^{-3}~M_\odot$. The average velocity of the ejected particles is $\approx 10^8$~cm~s$^{-1}$.

It is worth to mention that the above ejecta mass is also consistent with other independent merger simulations, e.g.~\citet{2014MNRAS.438...14D}, who showed that the amount of mass expelled in the merger can be obtained by the following fitting rational polynomial
\begin{equation}\label{eq:mej}
m_{\rm ej} = M \frac{0.0001807}{-0.01672 + 0.2463 q - 0.6982 q^2 + q^3},
\end{equation}
where $M = m_1+m_2$ is the total binary mass and $q \equiv m_2/m_1 \leq 1$ is the binary mass-ratio. Indeed, for the present case with $M = 1.4~M_\odot$ and $q=0.6/0.8$, the above formula gives $m_{\rm ej} = 0.00128~M_\odot$.

\begin{figure*}
\centering
\includegraphics[width=0.32\hsize,clip]{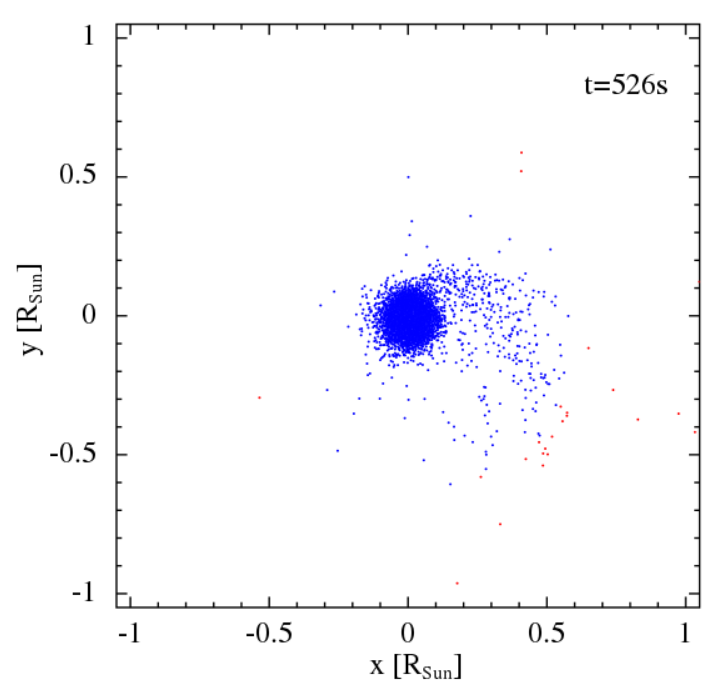}
\includegraphics[width=0.32\hsize,clip]{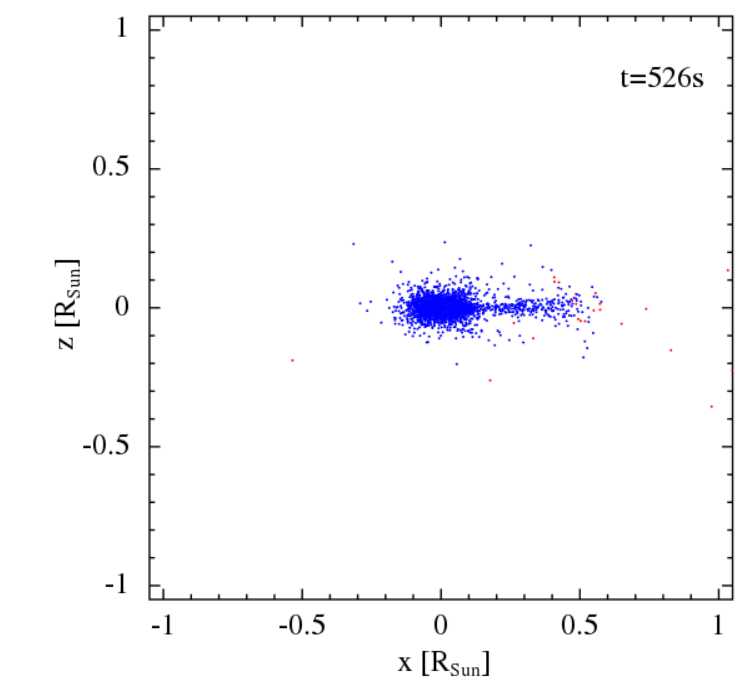}
\includegraphics[width=0.32\hsize,clip]{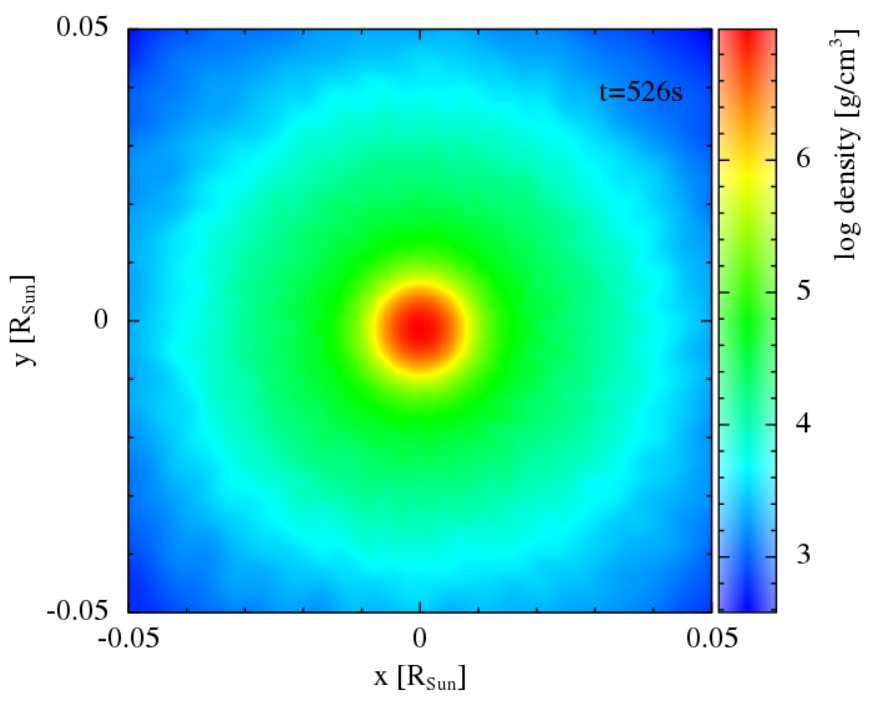}
\caption{Left panel: distribution of the SPH particles in the xy plane just after the merger. We can see a still dissipating spiral arm, the disk and the ejected particles. Bound particles are shown in blue and unbound particles are shown in red. Center panel: same as in the left panel but for the xz plane. Right panel: density (in g~cm$^{-3}$) plot in the xy plane. The central WD has a radius of $\approx 0.01~R_\odot$ while the disk is shown here up to $\approx 0.05~R_\odot$.}\label{fig:WDWDmerger2}
\end{figure*}

Figure~\ref{fig:WDWDmerger2} shows the distribution of the SPH particles in the xy and xz planes of the system as well as a density plot, just after the merger. It can be appreciated a still dissipating spiral arm, the disk and the ejected particles. We show the unbound particles in red and the bound particles in blue. It can be seen that the outer part of the spiral arm is gravitationally unbound while the inner region is bound and will fallback onto the newly-formed WD. With a mass of $1.1~M_\odot$ the central WD has a radius of $R_{\rm WD}\approx 5\times 10^8$~cm~$\lesssim 0.01~R_\odot$ \citep[see e.g.][]{2013ApJ...762..117B}, while the disk is shown here up to $\approx 0.05~R_\odot$.

\subsection{WD-WD merger rate}

The WD-WD merger rate has been recently estimated to be $(1$--$80)\times 10^{-13}$~yr$^{-1}$~$M_\odot^{-1}$ (at $2\sigma$) and $(5$--$9)\times 10^{-13}$~yr$^{-1}$~$M_\odot^{-1}$ (at $1\sigma$) \citep{2017MNRAS.467.1414M,2018MNRAS.476.2584M}. For a Milky Way stellar mass $6.4\times 10^{10}~M_\odot$ and using an extrapolating factor of Milky Way equivalent galaxies, 0.016~Mpc$^{-3}$ \citep{2001ApJ...556..340K}, it leads to a local cosmic rate $(0.74$--$5.94)\times 10^6$~Gpc$^{-3}$~yr$^{-1}$ ($2\sigma$) and $(3.7$--$6.7)\times 10^5$~Gpc$^{-3}$~yr$^{-1}$ ($1\sigma$). 

The above rate implies that $(12$--$22)\%$ of WD-WD mergers may end as type Ia SN. This is consistent with previously estimated rates of WD-WD mergers leading to SNe Ia \citep[see e.g.][]{2009ApJ...699.2026R}. We are here interested in the rest of the merger population not leading to Ia SNe.

\subsection{Magnetic field of the central WD}

The hot, rapidly rotating, convective corona can produce, via an efficient $\alpha\omega$ dynamo, magnetic fields of up to $B\approx 10^{10}$~G \citep[see e.g.][]{2012ApJ...749...25G}. Recent two-dimensional magneto-hydrodynamic simulations of post-merger systems confirm the growth of the WD magnetic field after the merger owing to the magneto-rotational instability \citep{Ji,2015ApJ...806L...1Z}. For a summary of the magnetic field configuration and its genesis in WD-WD mergers, as well as its role along with rotation in the aftermath of the dynamical mergers, see \citet{2018ApJ...857..134B}.

\section{Optical and infrared emission}\label{sec:3}

The ejected matter $m_{\rm ej}$ moves with an initial velocity $v_{{\rm ej},0}$ and we adopt for simplicity an evolving, uniform density profile
\begin{equation}\label{eq:rhoej}
\rho_{\rm ej} = \frac{3 m_{\rm ej}}{4 \pi r_{\rm ej}^3(t)},
\end{equation}
where $r_{\rm ej}(t)$ is the ejecta radius. 

The energy conservation equation is
\begin{equation}\label{eq:energybalance}
\frac{dE}{dt}=-P\frac{dV}{dt} - L_{\rm rad} + H,
\end{equation}
where $E$ is the energy, $P$ the pressure, $V = (4\pi/3) r_{\rm ej}^3$ is the volume, $L_{\rm rad}$ is the radiated energy and $H$ is the heating source, namely the power injected into the ejecta. For the ejecta we adopt a radiation dominated equation of state, namely $E = 3 P V$. The injected power $H$ is represented by the rotational energy coming from the spindown of the WD and the fallback accretion onto the WD:
\begin{equation}\label{eq:heating}
H = L_{\rm sd} + L_{\rm fb}.
\end{equation}

We adopt the spindown power by a dipole magnetic field 
\begin{equation}\label{eq:spindown}
L_{\rm sd} = \frac{2}{3} \frac{B_d^2 R^6}{c^3} \omega^4,
\end{equation}
where $\omega = 2\pi/P$ is the rotation angular velocity of the WD, $P$ is the rotation period, $B_d$ is the dipole field at the WD surface and $R$ is the WD radius. 

{Our assumption of the pulsar-like emission as part of the injection power into the ejecta is supported by the analysis of Sec.~\ref{sec:5.1} where we show that the magnetic field values of interest are larger than the minimum magnetic field needed for it not to be buried by the accreted matter. This indeed agrees with the recent results of \citet{2018ApJ...857..134B} on the thermal and rotational evolution of the central, massive WD produced by a WD-WD merger accounting for the torque by accretion, propeller, magnetic field-disk interaction and magneto-dipole emission. There, the timescale on which each of these regimes dominates the evolution has been obtained and it is shown the emergence of the magneto-dipole emission already at very early times post-merger even for the highest possible accretion rates which are higher than the ones considered in the present work.}

The fallback power can be parametrized by
\begin{equation}\label{eq:fallback}
L_{\rm fb} = L_{{\rm fb},0}\left(1+ \frac{t}{t_{\rm fb}} \right)^{-n},
\end{equation}
where $L_{{\rm fb},0}$ is the initial fallback luminosity, $t_{\rm fb}$ is the timescale on which the fallback power starts to follow a power-law behavior. {This function fits the numerical results by \citet{2009A&A...500.1193L} of the luminosity produced by the fallback of material of the disrupted secondary which remained bound in highly eccentric orbits. The derivation of this luminosity follows the treatment of \citet{2007MNRAS.376L..48R}. The material interacts with the disk in a timescale set by the distribution of eccentricities and not by viscous dissipation. The energy released is calculated as the difference in the kinetic plus potential energy of the particles between the initial position and the debris disk (dissipation) radius (obtained from the SPH simulation). Clearly, not all this energy can be released in form e.g. of photons to energize the ejecta so it has to be considered as an upper limit to the energy input from matter fallback.}

{Using Eq.~(\ref{eq:fallback})} we can also estimate the fallback accretion rate onto the WD as
\begin{equation}\label{eq:mdotfb}
\dot{m}_{\rm fb} \approx \frac{L_{\rm fb}}{G M_{\rm WD}/R_{\rm WD}}.
\end{equation}

Since little energy is radiated (see below) by the system, namely it is highly adiabatic, we can assume the radius to evolve according to \citep{2014MNRAS.439.3916M}
\begin{equation}\label{eq:vej}
\frac{1}{2} m_{\rm ej} v_{\rm ej}^2 \approx \frac{1}{2} m_{\rm ej} v_{{\rm ej},0}^2 + \int_0^{t} H dt,
\end{equation}
where $v_{\rm ej} \equiv d r_{\rm ej}/dt$ is the ejecta velocity. It is clear that in this most simple uniform density model under consideration this can be considered as a bulk average velocity. The density profile can have initially a radial dependence and in that case there would exist also a velocity profile with both faster and slower layers with respect to the unique one of our model.

Since the radiation travels on a photon diffusion timescale $t_{\rm ph} = r_{\rm ej}(1 + \tau_{\rm opt})/c$, the radiated luminosity can be written as
\begin{equation}\label{eq:Lrad}
L_{\rm rad} = \frac{c E}{r_{\rm ej}(1 + \tau_{\rm opt})},
\end{equation}
where
\begin{equation}\label{eq:opticaldepth}
\tau_{\rm opt} = \kappa \rho_{\rm ej} r_{\rm ej},
\end{equation}
is the optical depth with $\kappa$ the opacity. For the optical wavelengths and the composition of the merger ejecta we expect $\kappa \approx 0.1$--$0.2$~cm$^2$~g$^{-1}$. This is different from the higher opacity expected for r-process material composing the kilonova produced in NS-NS mergers.

The effective temperature of the observed blackbody radiation, $T_{\rm eff}$, can be obtained as usual from the bolometric luminosity equation
\begin{equation}\label{eq:Teff}
L_{\rm rad} = 4\pi r_{\rm ej}^2 \sigma T_{\rm eff}^4,
\end{equation}
where $\sigma$ is the Stefan-Boltzmann constant. {Being thermal, the density flux at the Earth from a source located at a distance $D$ is therefore}
\begin{equation}
    B_\nu = \frac{2\pi h \nu^3}{c^2} \frac{1}{\exp[h\nu/(k T_{\rm eff})] - 1} \left(\frac{r_{\rm ej}}{D}\right)^2,
\end{equation}
{where $\nu$ is the frequency.}

\begin{figure}
\centering
\includegraphics[width=\hsize,clip]{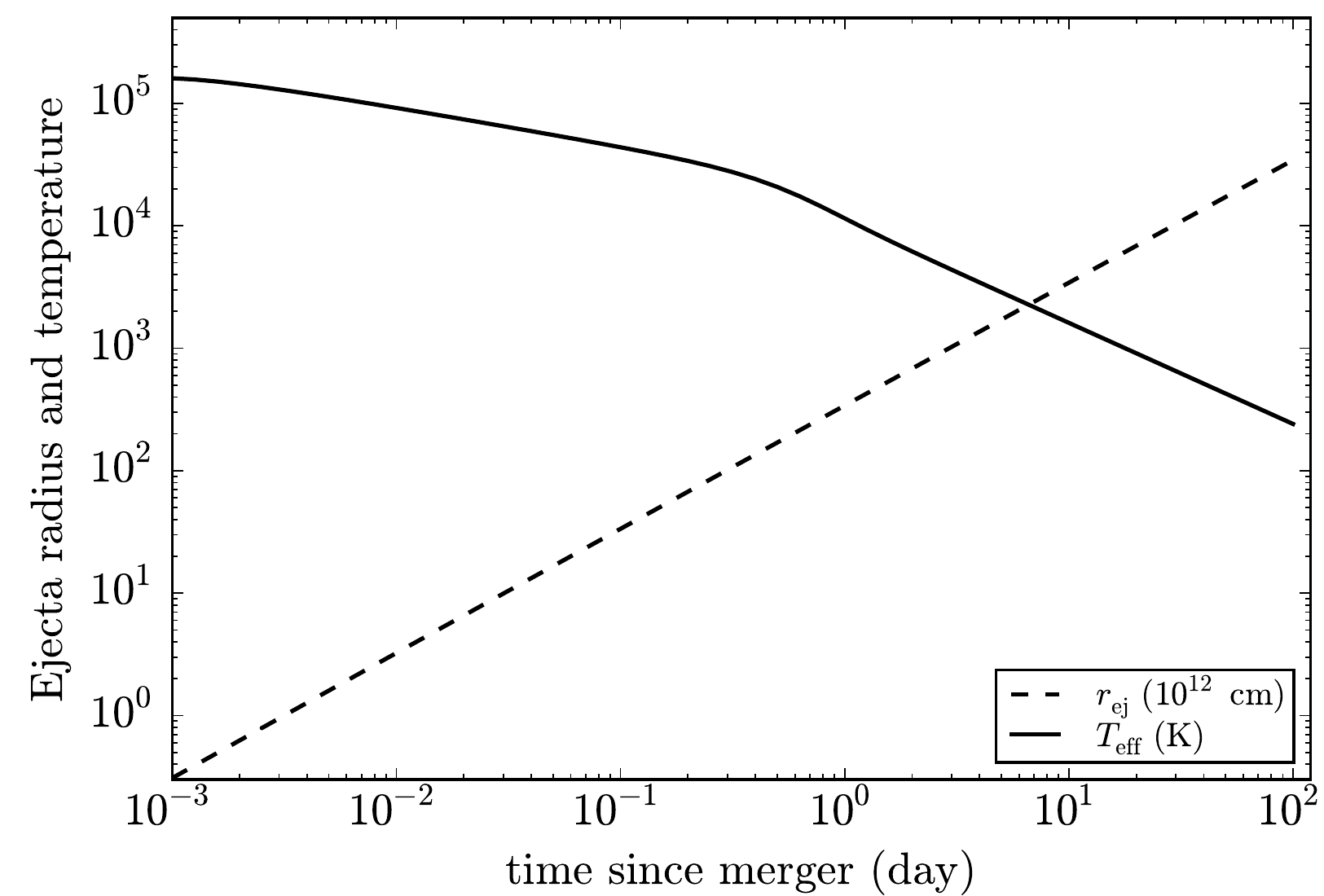}
\caption{
{Expected evolution of the ejecta radius and effective temperature from the cooling of $1.3\times 10^{-3}\,M_\odot$ ejecta heated by fallback accretion onto the newly-formed central WD.}
}\label{fig:RT}
\end{figure}

\begin{figure}
\centering
\includegraphics[width=0.49\hsize,clip]{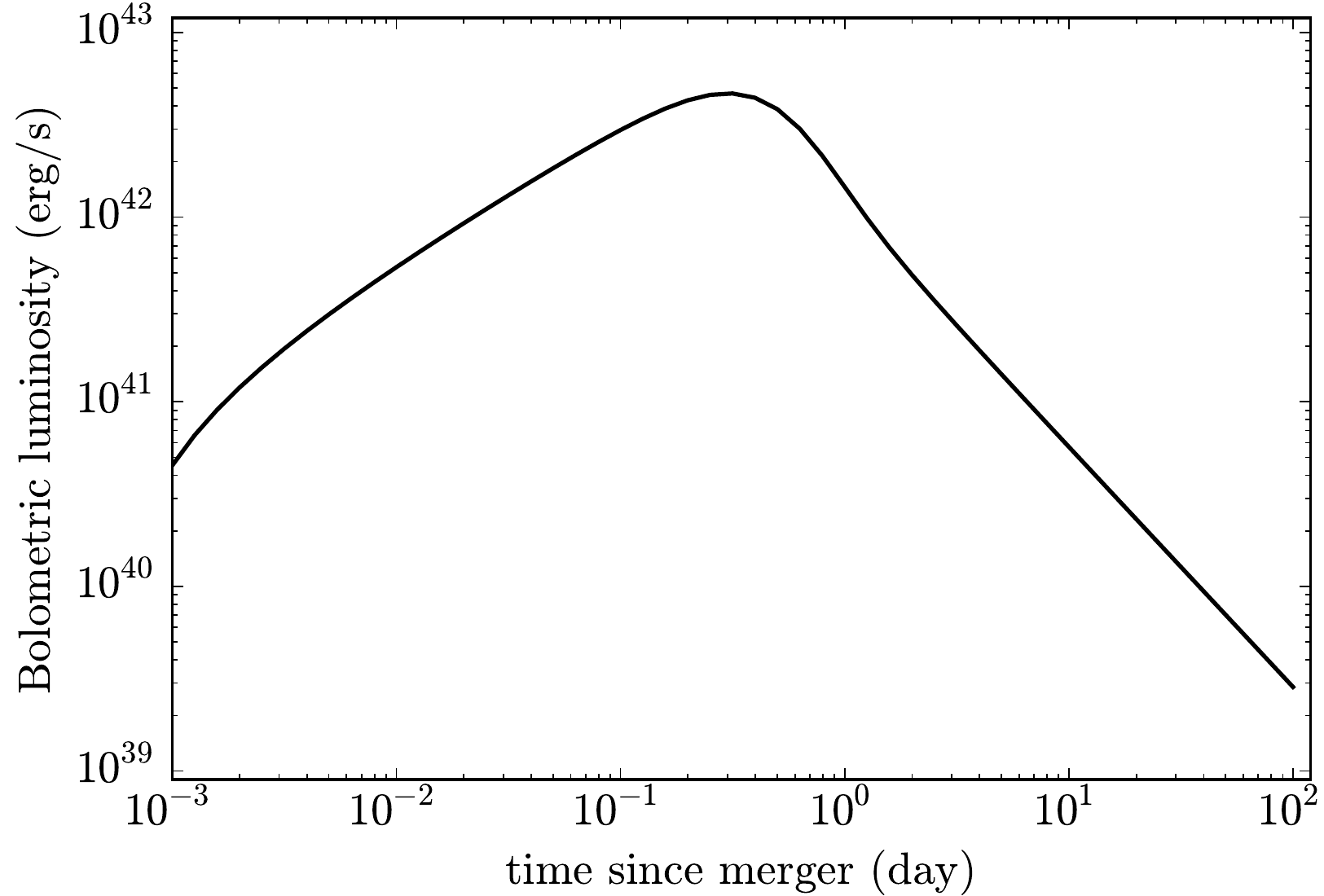}
\includegraphics[width=0.49\hsize,clip]{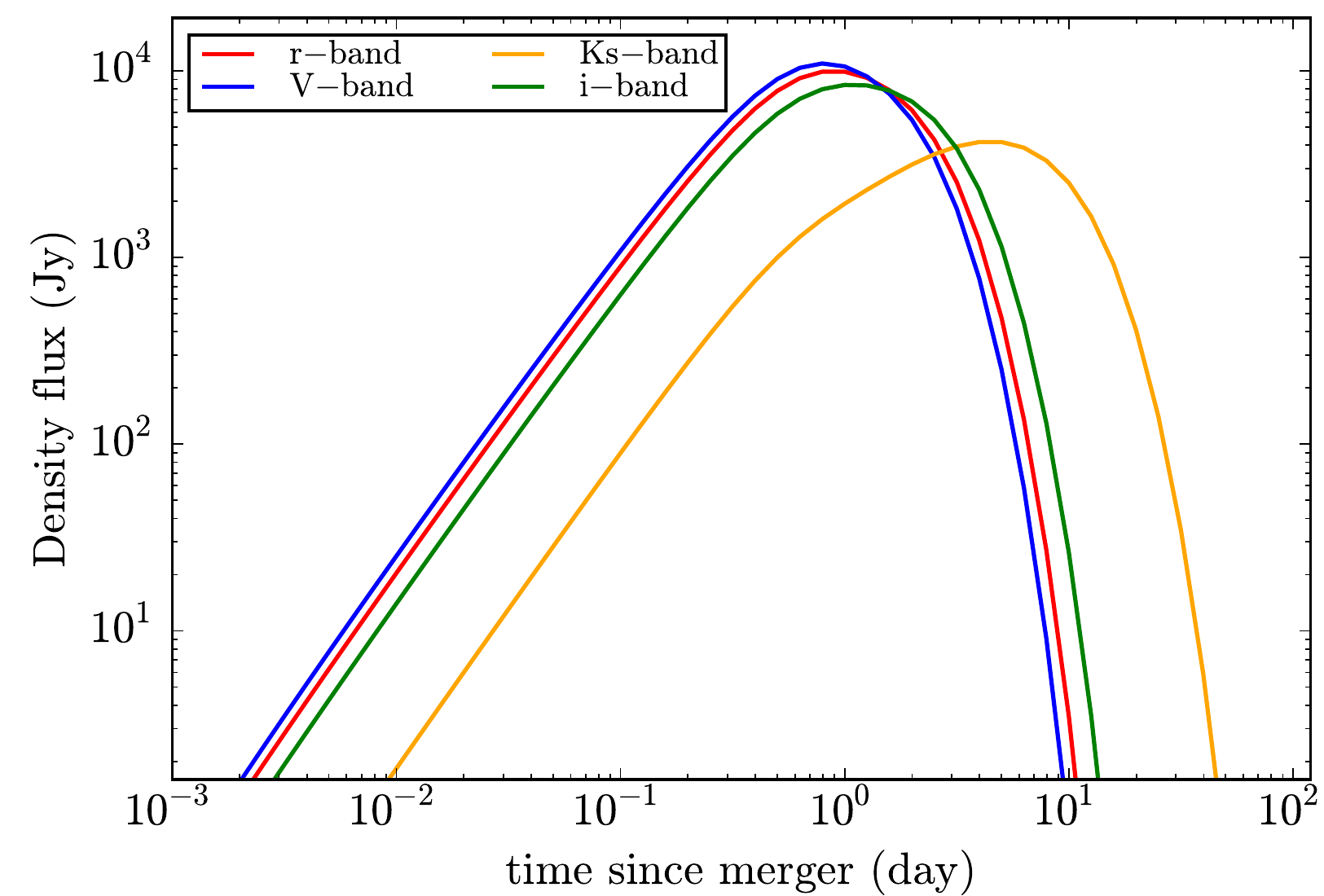}
\caption{
{Left panel: expected bolometric luminosity from the cooling of $1.3\times 10^{-3}\,M_\odot$ ejecta heated by fallback accretion onto the newly-formed central WD.
Right panel: corresponding expected flux density at Earth in the optical and in the infrared for a source located at $10$~kpc.}}\label{fig:wdwdejecta}
\end{figure}

{The ejecta radius $r_{\rm ej}$ and effective temperature $T_{\rm eff}$ obtained from the cooling of the merger ejecta are shown in Fig.~\ref{fig:RT}. Figure~\ref{fig:wdwdejecta} shows the expected bolometric luminosity (left panel) as well as the corresponding expected density flux at Earth (right panel) in the} optical and infrared{, for a source at $10$~kpc.}
%

We have chosen fallback power parameters according to numerical simulations of WD-WD mergers \citep[see e.g.~Sec.~5.3 and Fig.~8 in][]{2009A&A...500.1193L}: $L_{{\rm fb},0} = 8.0\times 10^{47}$~erg~s$^{-1}$ and $t_{\rm fb}=10$~s, and $n = 1.45$. For these parameters, it can be easily checked that the injection power from the WD spindown is negligible: even for a high field $B_d = 10^{10}$~G and an initial (at $t=0$) fast rotation period $P_0= 5$~s, we have $L_{\rm fb} = 8.0\times 10^{47}$~erg~s$^{-1}$ and $L_{\rm sd} = 9.6\times 10^{40}$~erg~s$^{-1}$, and for instance at $t = 1$~day, $L_{\rm fb} = 1.6\times 10^{42}$~erg~s$^{-1}$ and $L_{\rm sd} = 9.6\times 10^{40}$~erg~s$^{-1}$ (the spindown timescale for these WD parameters is much longer than one day). Thus, the ejecta is essentially only fallback-powered, namely $H\approx L_{\rm fb}$. 

Again, it is important to check the consistency of our fallback parameters with independent simulations. \citet{2014MNRAS.438...14D} showed that the fallback mass is well fitted by
\begin{equation}\label{eq:mfb}
m_{\rm fb} = M (0.07064 - 0.0648 q),
\end{equation}
which for our binary mass-ratio and total mass leads to $m_{\rm fb} = 0.031~M_\odot$. The fallback accretion leads to an energy injection 
\begin{equation}\label{eq:Efb}
E_{\rm fb} = \int L_{\rm fb} dt \approx \int\frac{G M_{\rm WD}}{R_{\rm WD}} \dot{m}_{\rm fb} c^2 dt \approx \frac{G M_{\rm WD}}{R_{\rm WD}} m_{\rm fb} c^2,
\end{equation}
where $m_{\rm fb}$ is the fallback mass given by Eq.~(\ref{eq:mfb}). For our present case, $M_{\rm WD}\approx 1.1~M_\odot$ and $R_{\rm WD}\approx 5\times 10^8$~cm, it leads to $1.79\times 10^{49}$~erg. This value has to be compared with the full integral $E_{\rm fb} = \int L_{\rm fb}dt \approx 1.78\times 10^{49}$~erg, where $L_{\rm fb}$ is given by Eq.~(\ref{eq:fallback}). The above estimate not only cross-checks the amount of fallback mass {as given independently by Eqs.~(\ref{eq:mdotfb}) and (\ref{eq:mfb})} but, at the same time, the mass and radius of the WD, as obtained from different simulations.

\begin{figure}
\centering
\includegraphics[width=\hsize,clip]{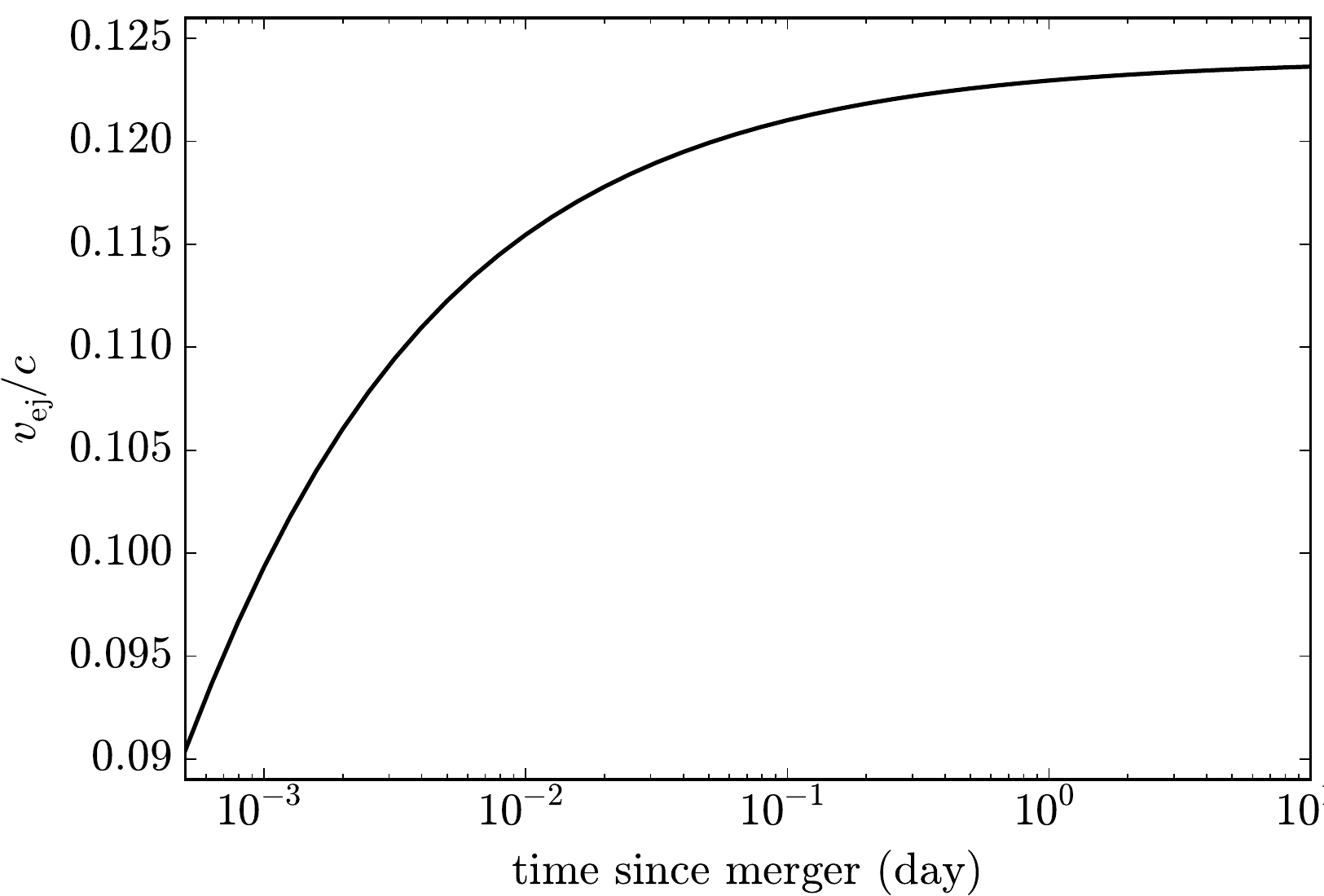}
\caption{Time-evolution of the ejecta velocity obtained from the integration of Eq.~(\ref{eq:vej}) accounting for the acceleration due to the presence of the heating source $H\approx L_{\rm fb}$.}\label{fig:vejecta}
\end{figure}

Since the ejecta are highly opaque at early times the fallback accretion and the spindown power are transformed into kinetic energy thereby increasing the expansion velocity of the ejecta; see Eq.~(\ref{eq:vej}). The matter is ejected a few orbits ($2$--$3$) before the merger and start to move outward with an initial non-relativistic velocity $0.01~c$ typical of the WD escape velocity. In the present example, such ejecta is then accelerated to mildly relativistic velocities $0.1~c$ (see Fig.~\ref{fig:vejecta}).

\section{X-ray emission}\label{sec:5}

The X-ray luminosity account for the absorption from the ejecta can be calculated as
\begin{equation}\label{eq:Lx}
L_X \approx \frac{1-e^{-\tau_X}}{\tau_X}(L_{\rm fb}+L_{\rm sd})\approx \frac{L_{\rm fb}+L_{\rm sd}}{1+\tau_X},
\end{equation}
where $\tau_X$ is the optical depth of the X-rays through the ejecta \citep[see e.g.][]{2018ApJ...856L..18M}:
\begin{equation}\label{eq:}
\tau_X = \kappa_X \rho_{\rm ej} r_{\rm ej},
\end{equation}
with $\kappa_X$ the opacity to the X-rays {which we assume to be dominated by bound-free electrons. For $1$--$10$~keV photons it can be in the range $10^2$--$10^4$~cm$^2$~g$^{-1}$ (see e.g.~Fig.~4 in \cite{2017LRR....20....3M} and \cite{1996ApJ...465..487V} for details), therefore for simplicity we here adopt a single value of $\kappa_X\approx 10^3$~cm$^2$~g$^{-1}$).}

In the above general discussion we have assumed that the WD can behave as a pulsar due to its dipole magnetic field and injects energy into the ejecta at a rate given by the radiation power given by Eq.~(\ref{eq:spindown}). However, we have first to check whether the magnetic field of the WD can be buried by the fallback accretion. 

\subsection{Is the magnetic field buried?}\label{sec:5.1}

The magnetic field is buried inside the star if the magnetospheric radius, 
\begin{equation}
R_m = \left(\frac{B^2 R_{\rm WD}^6}{\dot{m}_{\rm fb} \sqrt{2 G M_{\rm WD}}}\right)^{2/7},
\end{equation}
is smaller than the WD radius, $R_{\rm WD}$. Thus, using the value of $\dot{m}_{\rm fb}$ from Eq.~(\ref{eq:mdotfb}) we can compute the ratio $R_m/R_{\rm WD}$ and check if it is smaller or larger than unity. 

\begin{figure*}
\centering
\includegraphics[width=0.49\hsize,clip]{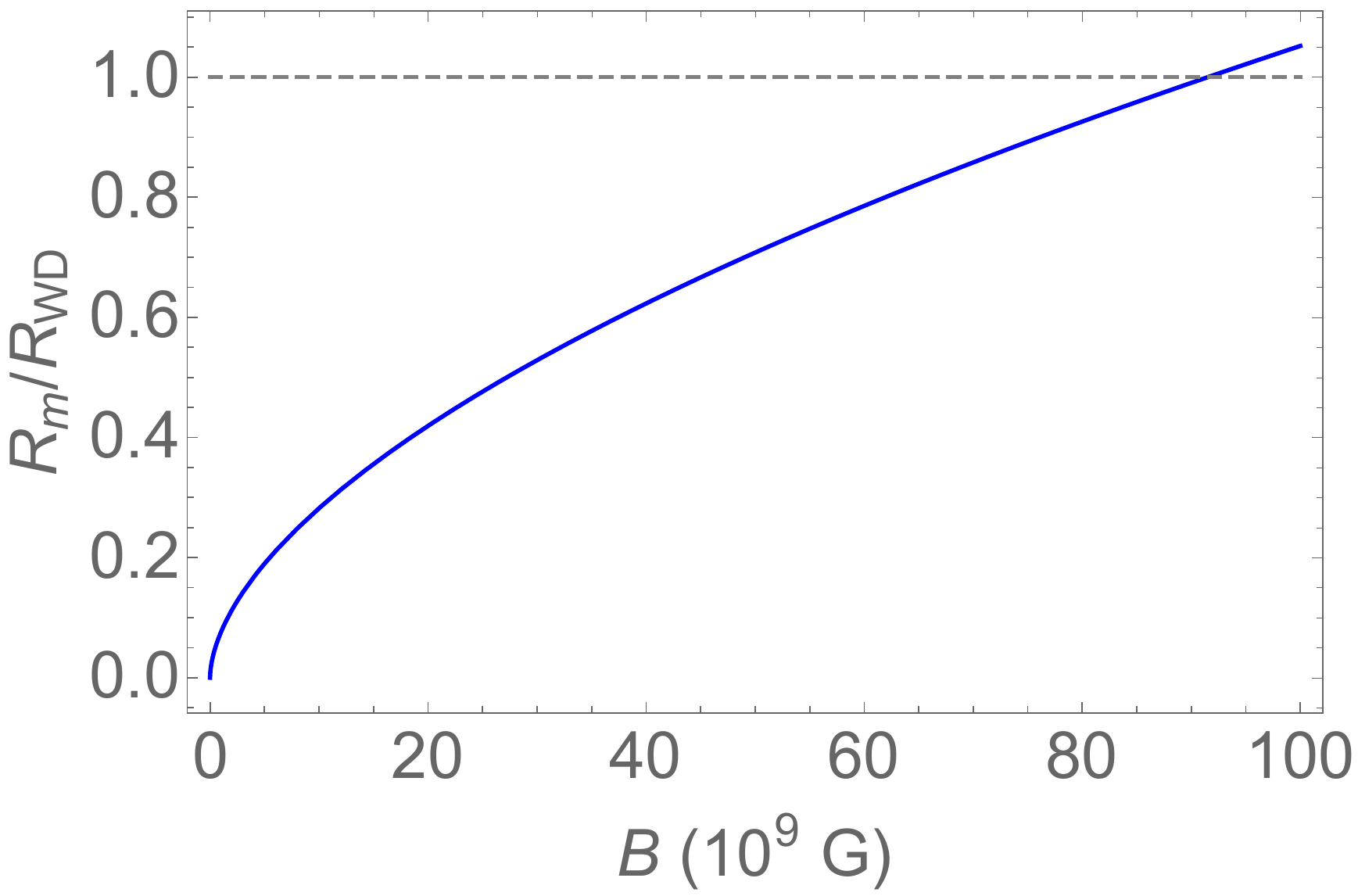}\includegraphics[width=0.49\hsize,clip]{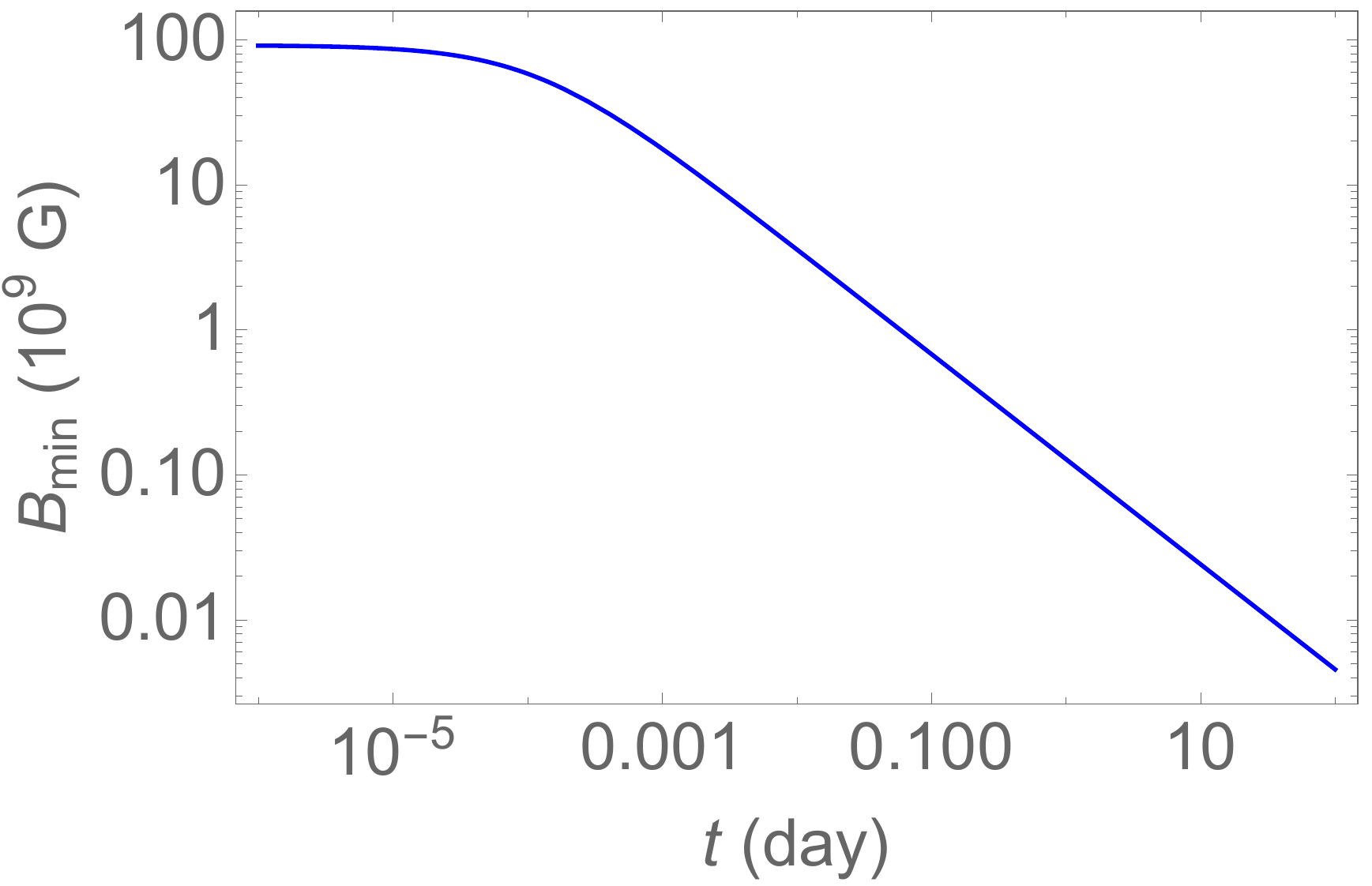}
\caption{Left panel: magnetospheric to WD radii ratio as a function of the WD surface magnetic field for an accretion rate set to the fallback value at time $t=0$ post-merger. Right panel: Minimum magnetic field $B_{\rm min}$ needed to have $R_m > R_{\rm WD}$, as a function of time, for the fallback accretion rate obtained from Eq.~(\ref{eq:mdotfb}). The mass and radius of the WD are $M_{\rm WD}=1.1 M_\odot$ and $R_{\rm WD}=5\times 10^8$~cm.}\label{fig:RmRcap}
\end{figure*}

The left panel of Fig.~\ref{fig:RmRcap} shows this ratio for an accretion rate set to the fallback value at time $t=0$ post-merger, while the right panel shows the value of the magnetic field for which $R_m = R_{\rm WD}$, say $B_{\rm min}$, as a function of time, for the fallback accretion rate given by Eq.~(\ref{eq:mdotfb}). $B_{\rm min}$ is the minimum value of the magnetic field that is not buried inside the star by the matter fallback. Therefore, for fields $B>B_{\rm min}$ the WD can behave as a pulsar and it can inject energy into the ejecta at the expenses of the WD rotational energy.

\subsection{Expected X-ray emission}\label{sec:5.2}

\begin{figure*}
\centering
\includegraphics[width=0.49\hsize,clip]{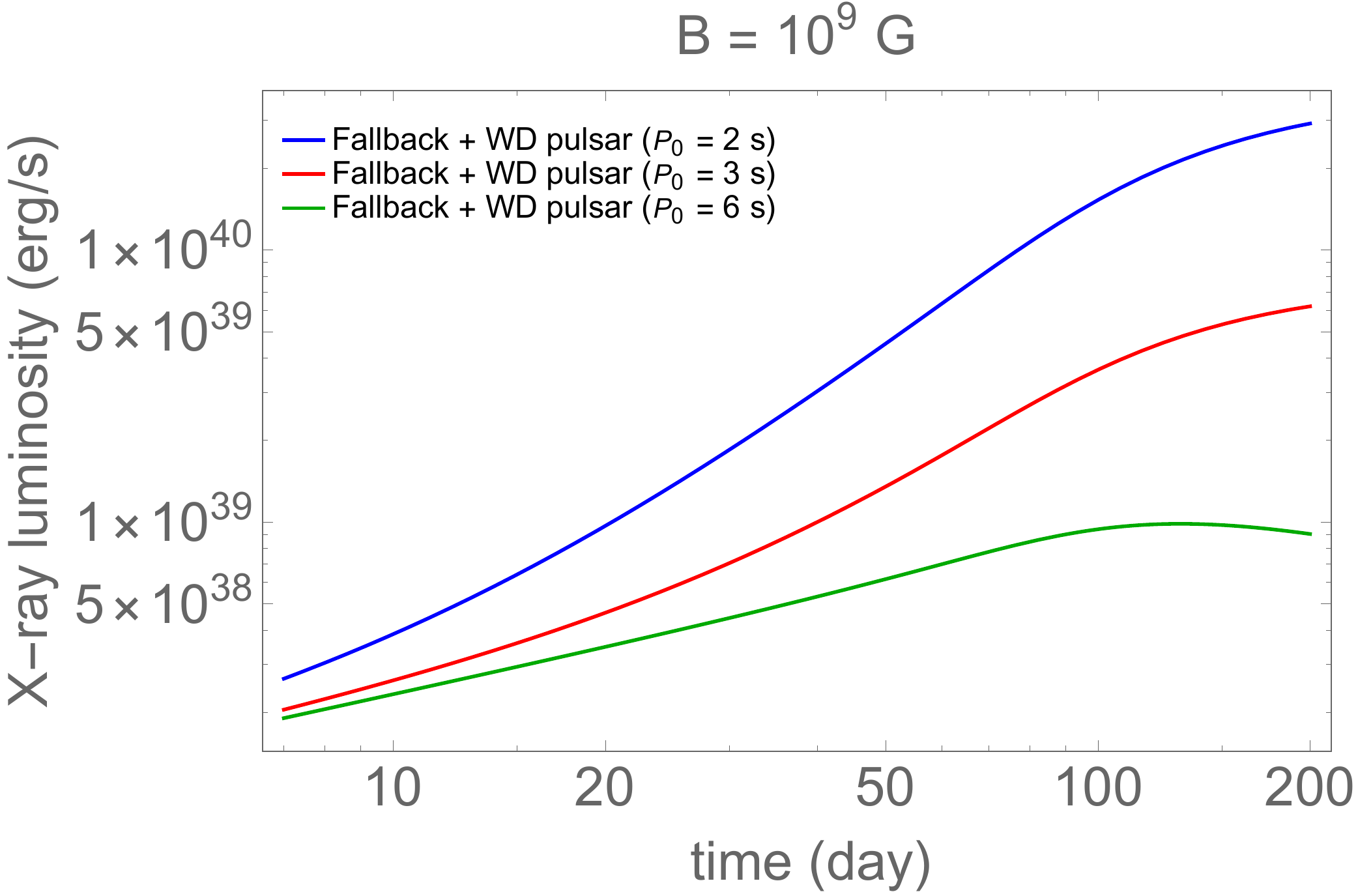}\includegraphics[width=0.49\hsize,clip]{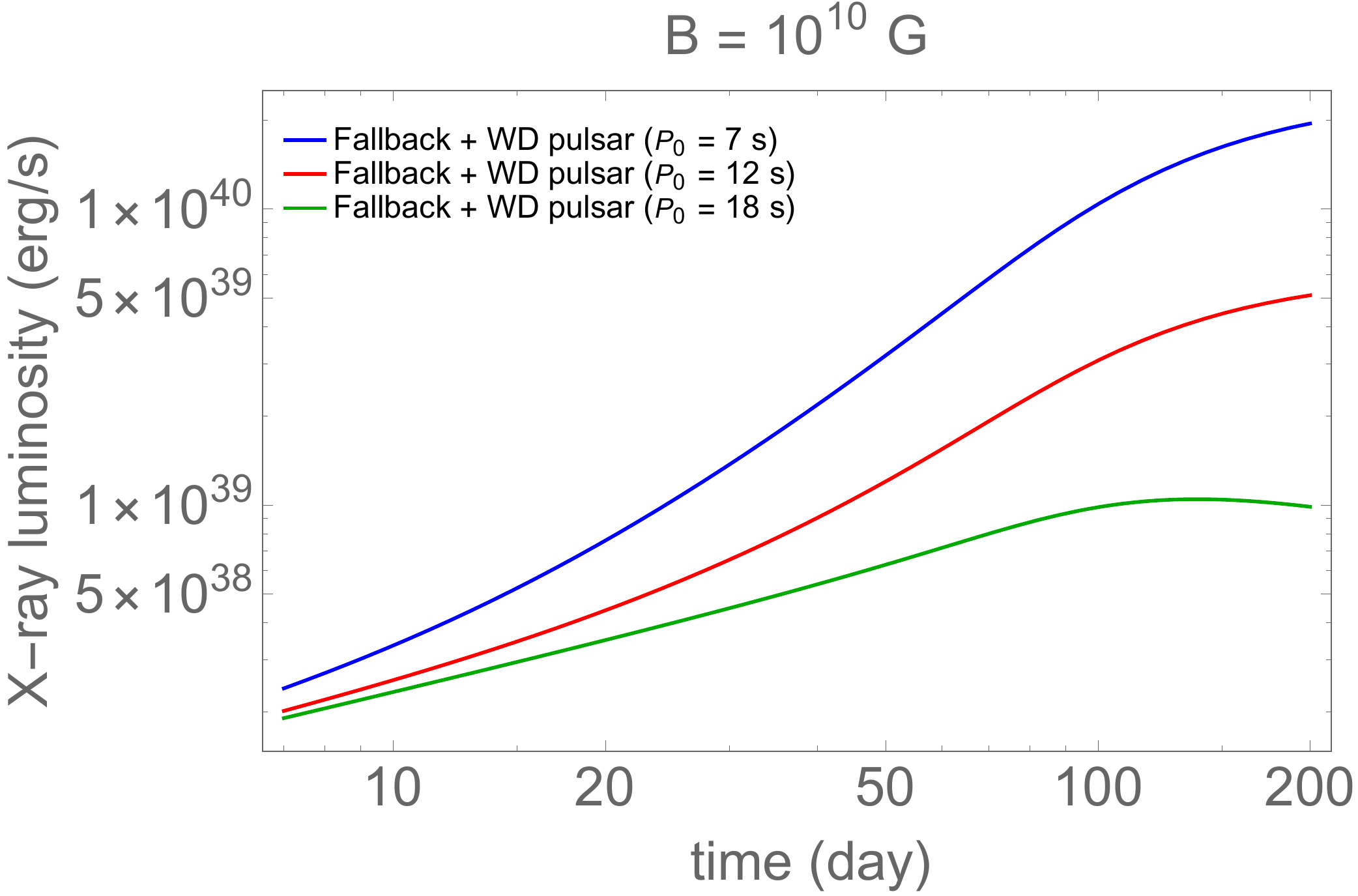}
\caption{Expected X-ray luminosity calculated via Eq.~(\ref{eq:Lx}) {for magnetic field values $10^9$~G (left panel) and $10^{10}$~G (right panel).}
}\label{fig:Lx}
\end{figure*}

Figure~\ref{fig:Lx} shows the X-ray luminosity (\ref{eq:Lx}) in comparison with the late-time X-ray emission data of GRB170817A. The comparison is made for selected values of the magnetic field, $B$, and the initial rotation period of the WD, $P_0$. 

It can be seen that a good agreement with the X-ray data can be obtained. Although the fallback power dominates over the pulsar one, the agreement is improved by adding the presence of the WD-pulsar (spindown) component. It is clear from our plots that the current X-ray data is not yet sufficient to unambiguously identify the WD parameters since an agreement is obtained for different combinations of $B$ and $P_0$. This is to be expected due to the magnetic dipole power dependence on the ratio $B^2/P^4$.

\subsection{WD-pulsar appearance}\label{sec:5.3}

From the above analysis we can see that additional data from other wavelengths, or X-ray data at later times, are needed to have an unambiguous identification of the WD parameters. Thus, it is interesting to compute when the newly-formed WD is expected to appear as a pulsar-like object in the sky. 

\begin{figure}
\centering
\includegraphics[width=\hsize,clip]{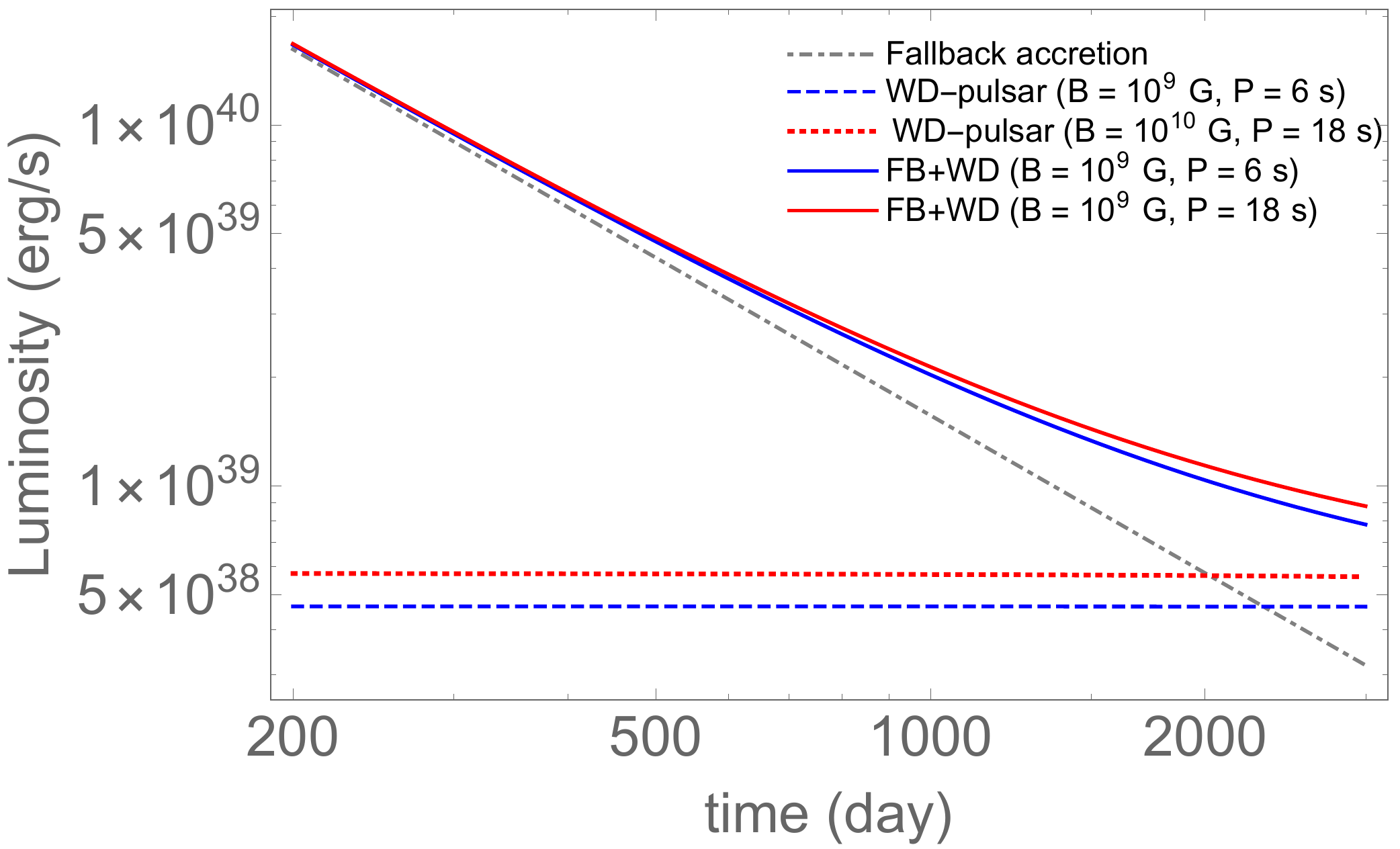}
\caption{Fallback versus spindown emission at times after the X-ray transparency. }\label{fig:fbsd}
\end{figure}

At the time of X-ray transparency, $t\sim 100$--$150$~day, the fallback power is still two orders of magnitude higher than the spindown one. However, at these post-merger timescales the fallback is fading continuously while the spindown power remains constant since, for the parameters in agreement with the current X-ray data (see Fig.~\ref{fig:Lx}), the spindown timescale is much longer. This implies that the WD can show up as a pulsar at a relatively early life of the post-merger system. To verify this we show in Fig.~\ref{fig:fbsd} the two components as a function of time after the X-ray transparency. At these times we can compare the unobserved luminosities given by Eq.~(\ref{eq:fallback}) and Eq.~(\ref{eq:spindown}) for the fallback and spindown power, respectively. It can be seen that in the two cases a deviation from the fallback power-law behavior to a less steep lightcurve decay appears at $t\gtrsim 500$~day. This is a predicted signature of the WD-pulsar presence. The precise crossing between the fallback power and the spindown component appears, in the case of $(B,P_0) = (10^9~{\rm G},6~{\rm s})$ and $(10^{10}~{\rm G},18~{\rm s})$, at $t=2318.3$~day ($6.3$~yr) and $2023.4$~day ($5.5$~yr), respectively.

\section{Gamma-ray emission}\label{sec:6}

The energy observed in gamma-rays in GRB 170817A, $E_{\rm iso}\approx 3 \times 10^{46}$~erg, can originate from flares owing to the twist and stress of the magnetic field lines during the merger process: a magnetic energy of $2\times 10^{46}$~erg is stored in a region of radius $10^9$~cm and magnetic field of $10^{10}$~G \citep{2012PASJ...64...56M}. Such a radius would imply a photon travel time of the order of $r/c \sim 0.1$~s, so a peak luminosity of few $10^{47}$~erg~s$^{-1}$.

We are also currently exploring the temperature properties of the ejecta at the beginning of the expansion. The ejected matter might have temperatures of the order of $10^8$~K at radii of $10^9$~cm which could clearly gives a luminosity of the order of $4\pi r^2 \sigma T^4 \approx 7\times 10^{46}$~erg~s$^{-1}$ with an energy peak of $\approx 3 k_B T\approx 30$~keV, so observable as a hard X-ray (soft gamma-ray) emission. If the matter expands adiabatically and isotropically then the temperature would decrease as $T\propto R^{-1}$ (adopting radiation-dominated matter) and therefore it can rapidly (in seconds timescale) fade to the soft X-rays to then become undetectable for the current X-ray satellites. The above makes the detection of this emission particularly difficult 
at early times post-merger.
These issues are important by their own and deserve further analysis in dedicated forthcoming works.

\section{Summary and conclusions}\label{sec:7}

{We have investigated the infrared, optical, X and gamma-ray emission associated with a WD-WD merger, the ejected matter and the post-merger signal from the newborn WD.}

In view of the high magnetic fields involved in the merger, a prompt emission in the gamma-rays 
might {occur as} the result of magnetospheric flaring activity owing to the twist and magnetic stresses. For instance, the release of magnetic energy associated with a field of $10^{10}$~G in a region of radius $10^9$~cm can lead to a total energy release of few $10^{46}$~erg in a burst of short ($\sim 0.1$--$1$~s) duration with a peak luminosity of few $10^{47}$~erg~s$^{-1}$.

We have modeled the time evolution of the ejecta as the expansion of a uniform density profile. 
{We show our results for a fiducial case of a $0.6+0.8~M_\odot$ WD-WD merger leading to a central WD of $1.1~M_\odot$ and ejecta mass $\sim 10^{-3}\,M_\odot$. The latter} start to move outward with initial bulk velocity $0.01~c$ and from a distance $\sim 10^9$~cm, typical of the escape velocity and radius from a WD-WD binary when the matter is ejected, i.e.~$2$--$3$ orbits before merger \cite{2009A&A...500.1193L}.

{The cooling of the expanding ejecta, heated by the fallback accretion onto the WD (see Sec.~\ref{sec:3}), results in a thermal emission observable in the infrared and in the optical. The bolometric luminosity associated with this thermal emission peaks with a value of $10^{41}$--$10^{42}$~erg~s$^{-1}$ about $0.5$--$1$~day post-merger (see Fig.~\ref{fig:wdwdejecta}).} We have shown that the ejecta initially expand at low, non-relativistic velocities $0.01~c$, to then being rapidly accelerated by the fallback energy injection to mildly-relativistic velocities of the order of $0.1~c$ (see Fig.~\ref{fig:vejecta}). 

The X-ray emission from the fallback accretion process (see Sec.~\ref{sec:5}) emerges and peaks with a value of $10^{38}$--$10^{39}$~erg~s$^{-1}$ at $100$--$150$~day post-merger (see Fig.~\ref{fig:Lx}). X-rays from the spindown power of the central WD become observable later at a time that depends on the WD parameters (see Fig.~\ref{fig:fbsd}).

{Once we have established for these systems their observable signatures across the electromagnetic spectrum and their nature, we can discuss some possibilities for their experimental identification.} We have shown that the mass, rotation period and magnetic field of the newly-formed central WD are similar to the ones proposed in the WD model of soft gamma-repeaters (SGRs) and anomalous X-ray pulsars (AXPs) \citep{2012PASJ...64...56M}. The merger rate is indeed enough to explain the Galactic population of SGRs/AXPs. Thus, if a WD-WD merger produced GRB 170817A-AT 2017gfo, an SGR/AXP (a WD-pulsar) may become observable in this sky position. As we have shown in Sec.~\ref{sec:6} (see Fig.~\ref{fig:fbsd}), the identification of first instants of the appearance of the WD-pulsar will allow to establish the WD parameters.

{In addition, it is remarkable that, as pointed out in \cite{2018JCAP...10..006R} and here further scrutinized, a WD-WD merger and the evolution of the ejecta powered by fallback accretion onto the newborn WD, is able to produce observational features in the X and in the gamma-rays similar to the ones of GRB 170817A and in the infrared and in the optical similar to the ones of AT 2017gfo.} 
The ejecta from a WD-WD merger are, nevertheless, different from the ejecta from a NS-NS merger in that: 1) they have a lighter nuclear composition and 2) they are powered by fallback accretion instead of the radioactive decay of r-process heavy nuclei. It is then clear that the spectroscopic identification of atomic species can discriminate between the two scenarios. However, such an identification has not been possible in any observed kilonovae since it needs accurate models of atomic spectra, nuclear reaction network, density profile, as well as radiative transport (opacity) which are not available at the moment.

\section*{Acknowledgements}
JMBL thanks support from the FPU fellowship by Ministerio de Educaci\'on Cultura y Deporte from Spain. JAR is grateful to Elena Pian for the interesting discussions on kilonovae.

\end{document}